# Doping optimization for the power factor of bipolar thermoelectric materials

Samuel Foster[*] and Neophytos Neophytou

School of Engineering, University of Warwick, Coventry, CV4 7AL, UK

[*]S.Foster@warwick.ac.uk

## Abstract

Bipolar carrier transport is often a limiting factor in the thermoelectric efficiency of narrow bandgap materials at high temperatures due to the reduction in the Seebeck coefficient and the introduction of an additional term to the thermal conductivity. Using the Boltzmann transport formalism and a two-band model, we simulate transport through bipolar systems and calculate their thermoelectric transport properties: the electrical conductivity, the Seebeck coefficient and the thermoelectric power factor. We present an investigation into the doping optimisation of such materials, showing the detrimental impact that rising temperatures have if the doping (and the Fermi level) is not optimised for each operating temperature. We also show that the doping levels for optimized power factors at a given operating temperature differ in bipolar systems compared to unipolar ones. We show finally that at 600 K, in a bipolar material with bandgap approximately that of $Bi_2Te_3$, the optimal doping required can reside between 10% - 30% larger than that required for an optimal unipolar material depending on the electronic scattering details of the material.

# I. Introduction

The efficiency of thermoelectric (TE) materials (which convert between heat and electricity) is quantified by the figure of merit $ZT = \sigma S^2 T/(\kappa_l + \kappa_e)$ where $\sigma$ is the electrical conductivity, $S$ is the Seebeck coefficient, $T$ is temperature, $\kappa_l$ is the lattice thermal conductivity and $\kappa_e$ is the electronic thermal conductivity. The quantity $\sigma S^2$ is termed the power factor (*PF*).

Many of the most important TE materials are narrow bandgap semiconductors.[1] These narrow bandgaps (e.g. PbTe ~ 0.3 eV,[2] $Bi_2Te_3$ ~ 0.2 eV,[3] SnSe ~ 0.39 eV)[4] mean the materials suffer from bipolar effects at high operating temperatures. The bipolar effect occurs when both electrons and holes contribute to charge transport. When this happens: i) $\kappa_e$ increases due to contributions from both electrons and holes, ii) an additional thermal conductivity term, the bipolar thermal conductivity, $\kappa_{bi}$, is introduced (a result of electron-hole recombination at the contacts),[5] which also introduces large increases in the Lorenz number,[6] iii) the Seebeck coefficient drops as both electrons and holes contribute to it with opposite signs, and iv) the Fermi level moves towards the midgap in order to conserve carrier concentration, (although it does not fall as quickly as in the unipolar case). The thermal conductivity from i) and ii) degrades thermoelectric performance through the denominator of *ZT*, whereas iii) degrades performance through the numerator.

The optimal thermoelectric performance (for both the peak *PF* and peak *ZT*) depends heavily on the carrier concentration,[7] and this optimal is known to be temperature dependent, i.e. the performance peaks at different doping concentrations for different temperatures.[8] However, although it is known that for unipolar materials the optimized doping increases as $T^{3/2}$,[9] the optimization of the carrier concentration for bipolar systems is not yet clarified.

While various strategies have been suggested to reduce the bipolar effect in order to regain high performance, such as using heterostructure designs,[10,11] band engineering to widen the bandgap,[12,13] grain boundaries with barriers for minority carriers,[14] in this work we show that considering proper doping optimization by taking into account the bipolar effects could also allow for performance improvements.

For this, in this work we use Boltzmann transport theory and a two-band model (conduction and valence band) to examine the impact of the bipolar effect on the thermoelectric transport coefficients ($\sigma$, $S$, and the *PF*), as well as its effect on the optimal carrier concentration and doping. We show that the typical models and trends employed in the literature for optimal doping concentrations for maximizing the power factor and *ZT* for a unipolar material are no longer valid in bipolar materials. We show that optimising the carrier concentration for the operating (higher) temperatures can provide significant increases in the power factor and *ZT* compared to maintaining a low temperature optimised carrier concentration.

## II. Approach

To calculate the thermoelectric coefficients we use the linearized Boltzmann transport formalism. In this method the electrical conductivity ($\sigma$), the Seebeck coefficient (*S*) and the electronic thermal conductivity ($\kappa_e$) are given by[15,16]

$$\sigma = q_0^2 \int_{-\infty}^{\infty} dE \left(-\frac{\partial f}{\partial E}\right) \Xi(E) \tag{1}$$

$$S = \frac{q_0 k_B}{\sigma} \int_{-\infty}^{\infty} dE \left(-\frac{\partial f}{\partial E}\right) \Xi(E) \left(\frac{E - E_F}{k_B T}\right) \tag{2}$$

$$\kappa_e = k_B^2 T \int_{-\infty}^{\infty} dE \left(-\frac{\partial f}{\partial E}\right) \Xi(E) \left(\frac{E - E_F}{k_B T}\right)^2 - \sigma S^2 T \tag{3}$$

where $q_0$ is the elementary charge, *E* is energy, *f* is the Fermi-Dirac distribution, $k_B$ is the Boltzmann constant, and $E_F$ is the Fermi level. The quantity $\Xi(E)$ is called the transport distribution function and is defined as

$$\Xi(E) = v^2(E) \tau(E) g(E) \tag{4}$$

where *v* is the bandstructure velocity, $\tau$ is the relaxation time and *g* is the density of states. Here we use the 3D density of states under an isotropic parabolic band approximation:

$$g(E) = \frac{m^{*3/2}}{\pi^2 \hbar^3} \sqrt{2(E - E_{C/V})} \tag{5}$$

where *m\** is the effective mass, $\hbar$ is the reduced Planck's constant, and $E_{C/V}$ is the conduction/valence band edge.

Acoustic phonon scattering (ADP) is considered under a relaxation time approximation, according to

$$\frac{1}{\tau} = \frac{\pi D_A^2 k_B T}{\hbar c_l} g(E) \tag{6}$$

where we use $D_A$ = 5 eV for the acoustic deformation potential as in typical semiconductors, and $c_l$=1.908×10$^{11}$ kgm$^{-1}$s$^{-2}$ is the elastic constant.[17]

Ionised impurity scattering (IIS) is included according to the Brooks-Herring model:

$$\tau = \frac{16\sqrt{2m^*}\pi\varepsilon_r^2\varepsilon_0^2}{N_I q^4}[\ln(1+\gamma^2) - \frac{\gamma^2}{1+\gamma^2}]^{-1} E^{3/2} \tag{7}$$

where $\varepsilon_r$ is the relative permittivity, $\varepsilon_0$ is the permittivity of free space, $N_I$ is the density of impurities and $\gamma^2 \equiv 8m^* E L_D^2/\hbar^2$ where

$$L_D = \sqrt{\frac{\varepsilon_r \varepsilon_0 k_B T}{q_0^2 N_I}} \tag{8}$$

is the Debye screening length.[17]

We consider two bandstructures as show in Fig. 1: i) a single parabolic conduction band with effective mass $m_C = m_0$ where $m_0$ is the electron rest mass, and conduction band edge $E_C = 0$ eV (unipolar case, Fig. 1 (a)); and ii) a bipolar system with a single parabolic conduction band with effective mass $m_C = m_0$ and $E_C = 0$ eV and a single parabolic valence band with effective mass $m_V = m_0$ and $E_V = -0.2$ eV (bipolar case, Fig. 1 (c)). The bandgap of the bipolar system ($E_g = 0.2$ eV) is similar to that of $Bi_2Te_3$, for example.

## III. Results

Most thermoelectric materials have complex bandstructures and even more complex scattering mechanisms, however, in this study we only employ the single band effective mass approximation, which can give us simple first order guidance towards doping optimization in bipolar materials, putting aside complexities that arise from multi-band features.

We begin by 'scanning' the Fermi level, $E_F$, across the unipolar and the bipolar bandstructure materials in order to identify the optimal values of the power factors and $ZT$ and the optimal positioning of the Fermi level (meaning that we compute the thermoelectric coefficients for a series of $E_F$ values, each $E_F$ corresponding to a specific doping concentration). We first consider the case in which transport is limited by acoustic phonon scattering (ADP) and then include ionised impurity scattering in addition (ADP+IIS). As we will show, the observations are different in the two cases.

In Fig. 2(a) and (b) we show the $PF$ versus $E_F$ for (a) the unipolar case, and (b) the bipolar case under ADP limited scattering at four different temperatures: $T = 300$ K (blue lines), $T = 400$ K (green lines), $T = 500$ K (red lines), $T = 600$ K (black lines). In the unipolar case it can be seen that the $PF$ peaks are just above the band edge (at approximately $E_C = 0$ eV) as previously suggested in earlier studies.[18-20] The Fermi level value at which this occurs increases linearly with temperature (a small shift only is evident here since the transition happens around 0 eV), but the peak *PF* remains constant. This behaviour will be discussed in more detail later. In the bipolar case (Fig. 2(b)) the *PF* peak for both bands moves even further into the band with increasing temperature. A small decrease is also unavoidable as the increasing contribution of holes from the valence band reduces *S*. Importantly, however, the *PF* peaks in both cases are spread over increasingly wider $E_F$ values with increasing temperature (the black lines are broader compared to the blue lines), meaning that the power factor is somewhat more resilient to changes in carrier concentration at higher temperatures. In Fig. 2(c) we show *ZT* versus $E_F$ for the bipolar case only (considering only $\kappa_e$, with $\kappa_l = 0$ for brevity, but which allows us to observe the peaks limiting case at very low $\kappa_l$ versus the limit of large $\kappa_l$, which follows the power factor trend). We do not show the unipolar case since, because $\kappa_e \propto \sigma$ in the non-degenerate limit, the quantity $ZT = \frac{\sigma S^2}{\kappa_e}$ diverges at low carrier concentrations, following the rise in *S*. In the bipolar case the peak occurs closer to the midgap than when the *PF* only is considered, although it also then rises more quickly with temperature as discussed later.

Although we considered ADP scattering alone, the high carrier concentration in TE materials is achieved by impurity doping, which introduces a strong, possibly dominant scattering mechanism in common semiconductors. Therefore, in Figs. 2(d)-(f) we further show the same three Fermi 'scans' in the presence of *both* acoustic phonon scattering and ionised impurity scattering (indicated as ADP+IIS). The introduction of an additional scattering mechanism reduces the power factor. However, as the temperature rises, in the ADP+IIS case the peak power factor value now increases with temperature in both the unipolar and bipolar cases. In the case of optimising *ZT*, the peaks again occur closer to the midgap (as in the ADP limited results), however the peak values are now higher in value. This is because, as seen in the transport distribution function shown in Fig. 1(d), the introduction of the IIS affects low energy electron more heavily than higher energy electrons. Since the Seebeck coefficient is proportional to the average energy of the current flow as $S \propto \langle E \rangle - E_F$ this results in an increase in the Seebeck coefficient (comparing at a fixed $E_F$). In addition, this also results in a widening of the 'effective transport bandgap' (although these states are available they contribute significantly less to transport). This then results in a decrease in the bipolar effect giving an additional increase to *S* as well as a reduction in $\kappa_e$. Hence the values of *ZT* increase with the addition of IIS.

To show the behaviour of the power factor as the temperature rises, we next take the bandstructures we consider with carrier concentration optimised at $T = 300$ K and examine how the thermoelectric coefficients change when that carrier concentration is kept fixed at the $T = 300$ K optimal value. This is in order to replicate the constant doping concentrations found in experimental set-ups. Figure 3 shows $\sigma$, $S$ and $PF$ versus $T$ for the unipolar (red lines) and bipolar (black lines) bandstructures for the cases of acoustic phonon scattering only (ADP, dashed lines) and acoustic phonon plus ionised impurity scattering (ADP+IIS, solid lines). Note that the optimal carrier concentration is different in the case of ADP and ADP+IIS situations, n = $3\times10^{19}$ cm$^{-3}$ and n = $6\times10^{19}$ cm$^{-3}$, respectively. As the temperature increases and the Fermi distribution broadens, $E_F$ drops in order to satisfy charge neutrality. The $E_F$ decrease is limited in the bipolar case due to the increasing contribution of holes to the total carrier concentration, which counteract the downshift of the $E_F$. The electrical conductivity decreases with temperature for two reasons: i) the acoustic phonon scattering strength is proportional to $T$ (see Eq. 6), ii) $E_F$ moves towards the midgap meaning lower velocity states are participating to transport. In the unipolar case (red lines), the Seebeck coefficient shows an increase with $T$ since it is proportional to the difference between the average energy of the current and the Fermi level, $S \propto \langle E \rangle - E_F$. The quantity $\langle E \rangle - E_F$ increases in the unipolar case due to the significant drop in $E_F$ with $T$. In the bipolar case, however, the Seebeck coefficient increases to a lesser extent compared to the unipolar case (and even eventually begins decreasing) due to the increase in holes which contribute to $S$ with opposite sign to the electrons. The resultant effect on the power factor from these behaviours is: i) in the unipolar ADP case (red dashed line), a decrease of ~15% from 300 K to 600 K is observed, ii) in the bipolar ADP case (black dashed line), despite the smaller reduction in $\sigma$ at 600 K from the extra contribution to current that the valence band provides, there is an overall degradation in the power factor by ~40% which is much more significant than in the unipolar case.

With the introduction of IIS for both unipolar and bipolar channels, $\sigma$ naturally drops due to the extra scattering rate. However, as expected, at higher temperatures this drop is not as substantial as in the ADP case as the IIS scattering typically weakens with temperature. This is due to the broadening of the Fermi distribution (see Fig. 1(b)) and the occupation of higher energy states with larger wavevectors which are less impacted by IIS. This can again be seen from the IIS stronger impact on the transport distribution function at lower energies in Fig. 1(d). The improvement in $\sigma$ from the valence band contribution in the ADP case in the bipolar channel (comparing red-dashed to black-dashed lines in Fig. 3(a)) is now also missing in the ADP+IIS lines due to the widening of the 'effective transport bandgap' that IIS causes as explained earlier, and effectively makes the material 'look' more unipolar (Fig. 2(d)).

When it comes to the Seebeck coefficient in Fig. 3(b) and the introduction of IIS, bipolar transport no longer has such a strong effect on $S$ with increasing temperatures, due to this widening of the 'effective transport bandgap' due to IIS, unlike in the ADP-limited case (black-solid versus black-dashed line in Fig. 3(b)). The result of these effects on the $PF$, therefore, is a significant reduction at low temperatures compared to the ADP-limited case, but an increase with temperature (Fig. 3(c)). The increase is a consequence of the smaller relative reduction in $\sigma$ and the continuous rising of $S$.

Figure 3 showed and explained why the power factor drops (in the ADP case) or increases less that its optimal value if the carrier concentration (controlled by doping) remains at the $T = 300$ K optimal levels. We now show that the power factor can be improved by a careful optimisation of the carrier concentration at higher temperature operations. In Fig. 4(a) we show the optimal $PF$ of the unipolar (red lines) and bipolar (black lines) bandstructures for the cases of ADP scattering only (dashed lines) and ADP+IIS (solid lines), i.e. the peaks of the Fermi scans seen in Fig. 2.

For ADP scattering only, whereas the unipolar system previously saw a reduction of ~15%, by optimising the doping with temperature the power factor now remains constant (Fig. 4(a) – red-dashed line). In the bipolar case the dramatic fall in the power factor due to the Seebeck reduction (as seen previously in Fig. 3(b), black-dashed line) is mitigated by increasing the Fermi level. Consequently the power factor, although still slightly decreasing with temperature, is now ~60% higher at 600 K than in the un-optimised case from Fig. 3(c) (un-optimised values from Fig. 3(c) shown by the square markers at 600 K in red (unipolar) and black (bipolar).

The Fermi level required to produce these optimal values rises linearly with temperature in the unipolar system (red-dashed line in Fig. 4(b)). This behaviour was earlier identified by Ioffe in Ref. 9 where it was shown that the optimal reduced Fermi level $\eta_{F,opt} = (E_F-E_C)/k_BT = r$, where $r$ is an exponent that depends on the electron scattering mechanism. Since $r$ is a constant, this gives $E_F \propto T$. In our case of acoustic phonon scattering $r = 0$, so we would expect the power factor to peak at the band edge. However, Ioffe's derivation assumes Boltzmann statistics for the carrier distribution and, indeed, running our calculations under that assumption reproduces such a result (not shown). However, using the more accurate (for degenerate doping conditions) Fermi-Dirac distribution, we find that in the case of acoustic phonon scattering $\eta_{F,opt} \approx 2/3$. In the bipolar system the linear behaviour seen in the unipolar case no longer holds, and the optimum Fermi level rises quicker than linearly (black-dashed line in Fig. 3(b)). This is in order to avoid the detrimental impact of the bipolar effect that the valence band introduces.

In Fig. 4(c) we also show the optimal carrier concentration required to set $E_F$ at the optimal position. As has been previously identified in the literature,[9,21] the optimal carrier concentration in a unipolar system increases as $n_{opt} \propto T^{3/2}$ (red-dashed line). Again, however, in the bipolar system (black-dashed line) the unipolar behaviour no longer holds, and the required carrier concentration rises more quickly in order to produce the higher Fermi levels seen in Fig. 4(b), following an approximate $T^{1.8}$ trend. Indeed, at $T = 600$ K the optimal bipolar carrier concentration is 30% higher than the optimal unipolar carrier concentration.

When IIS is included, the power factor values are lower as explained previously, but increase with increasing temperature due to the occupation of higher energy states which scatter less under IIS. Benefits compared to the un-optimised values (diamond markers in Fig. 4(a)) are not as great as in the ADP only case, but still significant – 10% for the unipolar bandstructure and 20% for the bipolar bandstructure (solid lines in Fig. 4(a)). The Fermi level and carrier concentration values needed to achieve these power factor values are higher than in the ADP only case. For practical purposes, therefore, to achieve an optimized power factor in the bipolar case at $T = 600$ K in the material we consider of bandgap $E_g$=0.2 eV, the doping concentration needs to be by 160% higher compared to the value that provides optimized $PF$ at $T = 300$K. That value is by 10% higher compared to the one that achieves the optimal $T = 600$ K $PF$ in the unipolar case. Note that in the case of the ADP+IIS transport conditions, the optimal doping density is higher, due again to the widening of the 'effective transport bandgap' seen in Fig. 1(d). Also note that these values are to be altered in the case of a different bandgap, i.e. the relevance of these values are shifted to lower/higher temperatures as the bandgap decreases/increases.

Finally, due to the influence of the thermal conductivity in the denominator of $ZT$, which has its own temperature dependence, $ZT$ does not peak at the same $E_F$ or carrier concentration as the $PF$. Therefore, in Fig. 5 we compare the optimal carrier concentration and Fermi levels when optimising for the power factor (same black lines as in Fig. 4 (b) and (c)) and optimising for $ZT$ (green lines). This comparison here is shown only for the bipolar material since the unipolar material does not show a peak as explained previously. For the calculation of $ZT$ we consider only the electronic properties (i.e. we take $\kappa_l = 0$, as the behaviour of $\kappa_l$ is material dependent and more complex). Since $\kappa_e \propto \sigma$ through the Lorenz number, $\kappa_e$ is reduced with falling $E_F$ and, therefore, the peaks in $ZT$ occur at significantly lower density and $E_F$ than when just optimising for the $PF$. As the temperature is increased, however, the optimal values (in both ADP and ADP+IIS cases) rise at a quicker pace than when optimising for $PF$. This is because as the temperature increases the impact of the bipolar effect kicks in and $\kappa_{bi}$ increasingly pushes the peak away from the midgap. The introduction of IIS, however, when optimising for $ZT$ has much less influence than in when optimising for the $PF$. This is again due to $\kappa_e$ being proportional to $\sigma$. As can be seen in Fig. 3, the introduction of IIS primarily affects $\sigma$. When optimising for $ZT$, this impact is then cancelled out by the same impact on $\kappa_e$.

Of course in a real material $\kappa_l \neq 0$ and the optimal $ZT$ values will lie somewhere between the $PF$-optimised and our $\kappa_l = 0$ $ZT$-optimised values. In particular, it is interesting to note that the smaller the value of $\kappa_l$ in the material with respect to the $\kappa_e$, the closer it is to the $\kappa_l = 0$ $ZT$-optimised case, and therefore the less it needs to be doped to reach its optimal $ZT$, which can prove helpful for TE materials, as doping at extremely high values can prove difficult in many cases.

Finally, we would like to state that in this work we employed a simple two-band parabolic model to obtain first order optimization strategies for doping in bipolar TE materials. In reality, material bandstructures are typically more complex than the simple two-band parabolic model we assume here. Real material bandstructures can have a variety of band gaps, effective masses, band degeneracies, band non-parabolicity, and multiple valence and/or conduction bands. Many of these bandstructure features can also vary with temperature, and detailed studies on each material are essential for proper optimization. In this study however, it was our aim to demonstrate to first order the important, yet overlooked, impact of the bipolar effect on the doping optimisation.

## IV. Conclusions

Using the Boltzmann transport formalism we have calculated the thermoelectric transport coefficients for unipolar and bipolar systems and presented a study on the optimal doping conditions for the power factor and $ZT$ figure of merit. We have shown that if the carrier concentration is not properly optimised at the temperature of operation, but room temperature optimal doping is considered, the power factor can underperform by 15% in the unipolar systems, and 40% in the bipolar system under ADP scattering, and 10% in the unipolar systems, and 20% in the bipolar system under ADP+IIS scattering. Consequently, significant enhancements in the *PF* (~40%) can be achieved through doping optimisation. Furthermore we have identified that in a bipolar system the optimal carrier concentration indicates an approximately $T^{1.8}$ trend, larger compared to the $T^{3/2}$ trend in unipolar materials, a result of the additional degradation due to bipolar transport. In our simulations, the optimal carrier concentration at $T$ = 600 K in a material with bandgap $E_g$ = 0.2 eV (e.g. approximately that of $Bi_2Te_3$) then becomes 30% larger than expected from the unipolar calculation. We believe that our findings will be useful in the optimal design of bipolar thermoelectric materials.

Acknowledgements: This work has received funding from the European Research Council (ERC) under the European Union's Horizon 2020 Research and Innovation Programme (Grant Agreement No. 678763).

Figure 1:

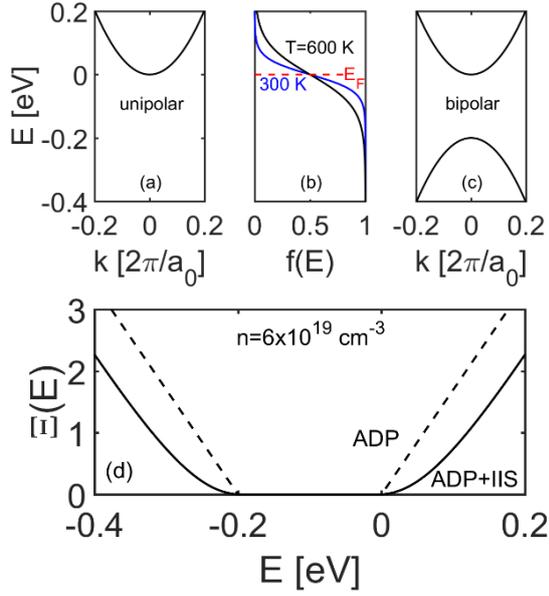

## Figure 1 caption:

(a) The unipolar case: a single parabolic conduction band with effective mass $m_C = m_0$ and conduction band edge $E_C = 0$ eV, (b) the Fermi distribution at $T = 300$ K (blue line) and $T = 600$ K (black line) with $E_F = 0$ eV (red-dashed line), and (c) the bipolar case: a single parabolic conduction band with effective mass $m_C = m_0$ and $E_C = 0$ eV and a single parabolic valence band with effective mass $m_V = m_0$ and $E_V = -0.2$ eV. In (d) we show the transport distribution function verses energy for the bipolar material for two different scattering regimes: acoustic phonon scattering (dashed line), and acoustic phonon scattering and ionised impurity scattering for an impurity density of $n = 6\times10^{19}$ cm$^{-3}$ (solid line).

Figure 2:

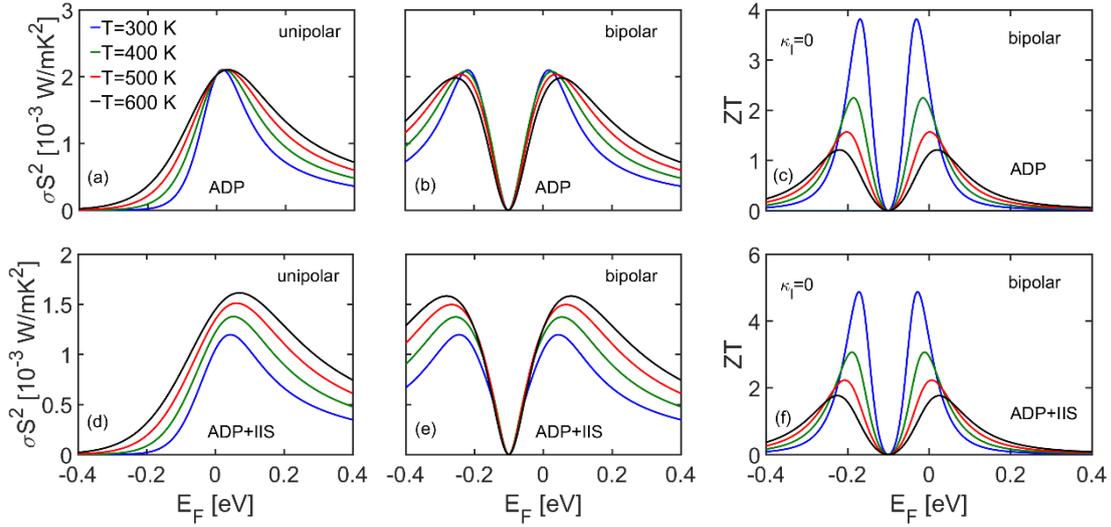

Figure 2 caption:

The power factor versus Fermi level at four different temperatures: 300 K (blue lines), 400 K (green lines), 500 K (red lines), 600K (black lines) for (a) a single parabolic conduction band with $E_C = 0$ eV and $m_C = m_0$, under acoustic phonon scattering conditions (ADP), (b) a bipolar system with one parabolic conduction band with $E_C = 0$ eV and $m_C = m_0$, and one parabolic valence band with $E_V = -0.2$ eV and $m_V = m_0$ under acoustic phonon scattering conditions, (d) a single parabolic conduction band with $E_C = 0$ eV and $m_C = m_0$, under acoustic phonon and ionised impurity scattering conditions (ADP+IIS), and (e) a bipolar system with one parabolic conduction band with $E_C = 0$ eV and $m_C = m_0$, and one parabolic valence band with $E_V = -0.2$ eV and $m_V = m_0$ under acoustic phonon and ionised impurity scattering conditions. In (c) and (f) we show $ZT$ (with $\kappa_l = 0$) versus Fermi level for the same four temperatures, and for ADP, and ADP+IIS conditions, respectively.

Figure 3:

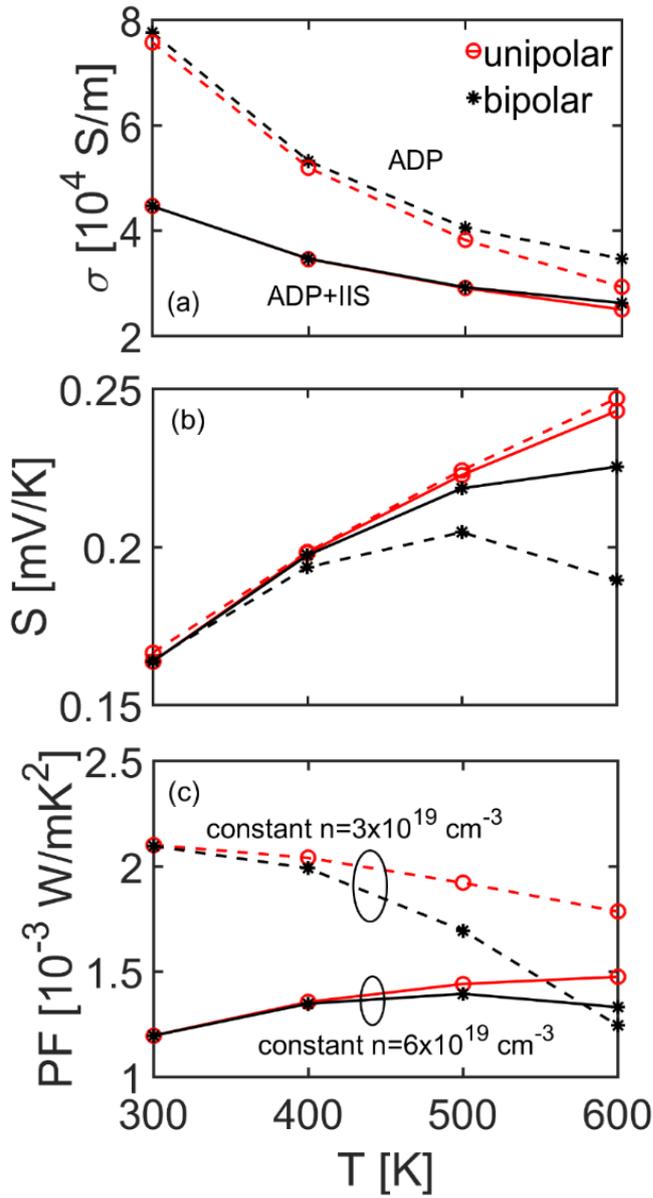

## Figure 3 caption

The (a) electrical conductivity, (b) Seebeck coefficient, and (c) power factor versus temperature at constant carrier concentration for two bandstructures: a single parabolic conduction band of mass $m_c = m_0$, (red lines), and a bipolar system with one conduction and one valence band with masses $m_c = m_0$, $m_v = m_0$ (black lines), and bandgap $E_g = 0.2$ eV. Results are shown for acoustic phonon scattering, ADP, only (dashed lines), and for acoustic phonon and ionised impurity scattering, ADP+IIS (solid lines).

Figure 4:

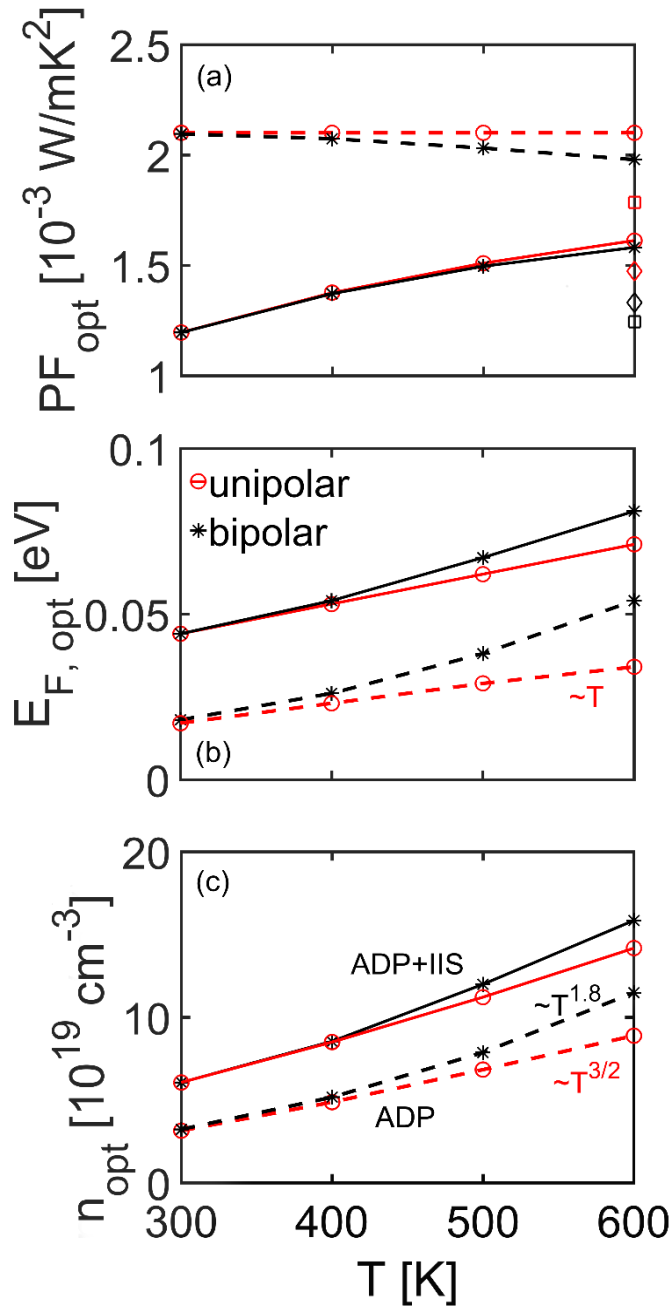

Figure 4 caption:

The optimal values of (a) the power factor, (b) Fermi level, and (c) carrier concentration versus temperature for two bandstructures: a single parabolic conduction band of mass $m_c = m_0$, (red lines), and a bipolar system with one conduction and one valence band with masses $m_c = m_0$, $m_v = m_0$ (black lines), and bandgap $E_g = 0.2$ eV. Results are shown for acoustic phonon scattering only, ADP (dashed lines), and for acoustic phonon and ionised impurity scattering, ADP+IIS (solid lines).

Figure 5:

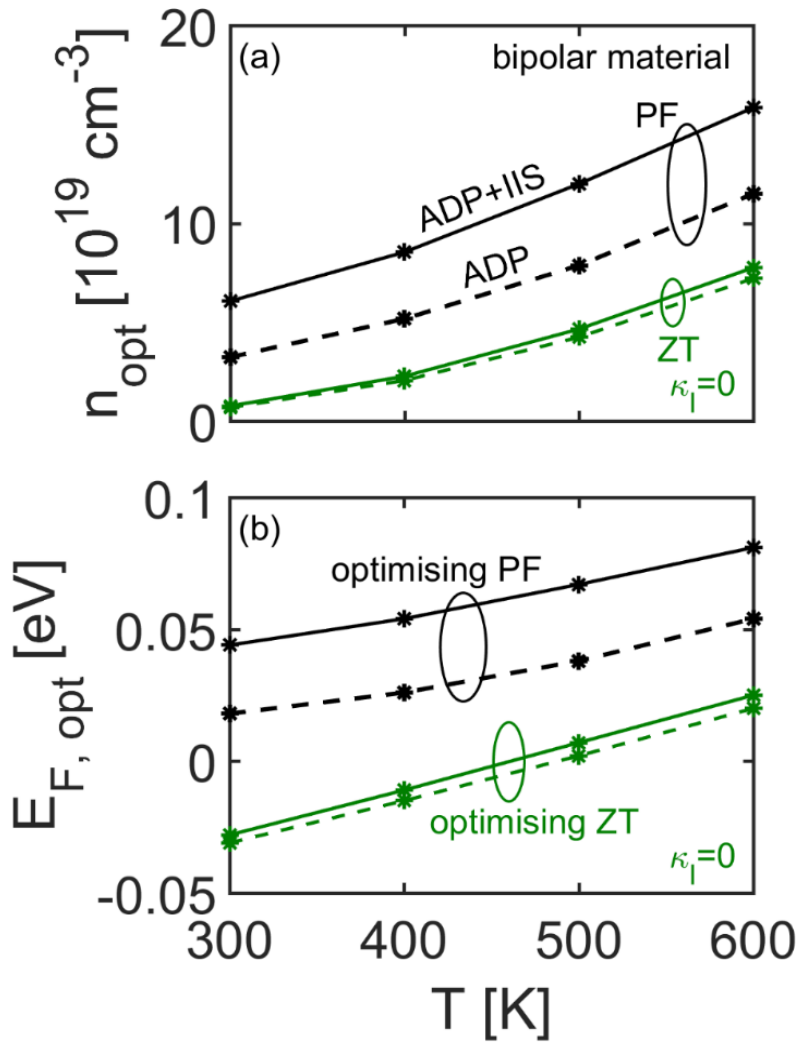

Figure 5 caption:

The optimal values of (a) carrier concentration and (b) the Fermi level versus temperature for a bipolar bandstructure to maximise the power factor (black lines) and *ZT* (green lines) in the case of acoustic phonon scattering only, ADP (dashed lines) and acoustic phonon and ionised impurity scattering, ADP+IIS (solid lines).